\def\mytitle   {Convex Hulls}
\def\mysubtitle{Surface Mapping onto a Sphere}
\def\myauthor  {Ben Kenwright} 
\def\myemail   {bkenwright@ieee.org (Dec 2014)} 
\def\mykeywords{convex hulls, 3d, 2d, iterative, stable, coplanar, co-linear, reliable, engineering challenges, support mapping, poly, convex, concave, computer generated, interactive}

\documentclass[conference,backref=page]{acmsiggraph}

\TOGonlineid{45678}
\TOGvolume{0}
\TOGnumber{0}
\TOGarticleDOI{1111111.2222222}
\TOGprojectURL{}
\TOGvideoURL{}
\TOGdataURL{}
\TOGcodeURL{}

\title{\mytitle \\ \fontsize{13}{16}\selectfont \mysubtitle}
\author{\myauthor\thanks{e-mail:\myemail} }
\pdfauthor{\myauthor}

\usepackage[compact]{titlesec}
\titlespacing{\section}{0pt}{0ex}{0ex}
\titlespacing{\subsection}{0pt}{0ex}{0ex}
\titlespacing{\subsubsection}{0pt}{0.0ex}{0ex}

\hypersetup{
	colorlinks  = true, 
	linkcolor   = blue,
	anchorcolor = red,
	citecolor   = blue, 
	filecolor   = red, 
	pdftex,
	pdfauthor={\myauthor},
	pdftitle={\mytitle},
	pdfsubject={\mytitle},
	pdfkeywords={\mykeywords},
	pdfcreator={\myauthor},
	pdfproducer={\myauthor},
}

\usepackage{enumitem}
\setitemize{leftmargin=13.0pt,noitemsep,topsep=0pt,parsep=0pt,partopsep=0pt}
\setenumerate{noitemsep,topsep=0pt,parsep=0pt,partopsep=0pt}

\usepackage{graphicx}

\graphicspath{{./images/}}

\newcommand{\figuremacroW}[4]{
	\begin{figure}[!htbp] 
		\centering
		\includegraphics[width=#4\columnwidth]{#1}
		\caption[#2]{\textbf{#2} - #3}
		\label{fig:#1}
	\end{figure}
}

\newcommand{\figuremacroF}[4]{
	\begin{figure*}[!htbp] 
		\centering
		\includegraphics[width=#4\textwidth]{#1}
		\caption[#2]{\textbf{#2} - #3}
		\label{fig:#1}
	\end{figure*}
}

\newcommand{\figuremacroFb}[4]{
	\begin{figure*}[!btp] 
		\centering
		\includegraphics[width=#4\textwidth]{#1}
		\caption[#2]{\textbf{#2} - #3}
		\label{fig:#1}
	\end{figure*}
}

\usepackage{amssymb}

\usepackage[ruled,linesnumbered]{algorithm2e}

\usepackage{setspace}

\begin{document}


\teaser{
 \includegraphics[width=1.0\textwidth]{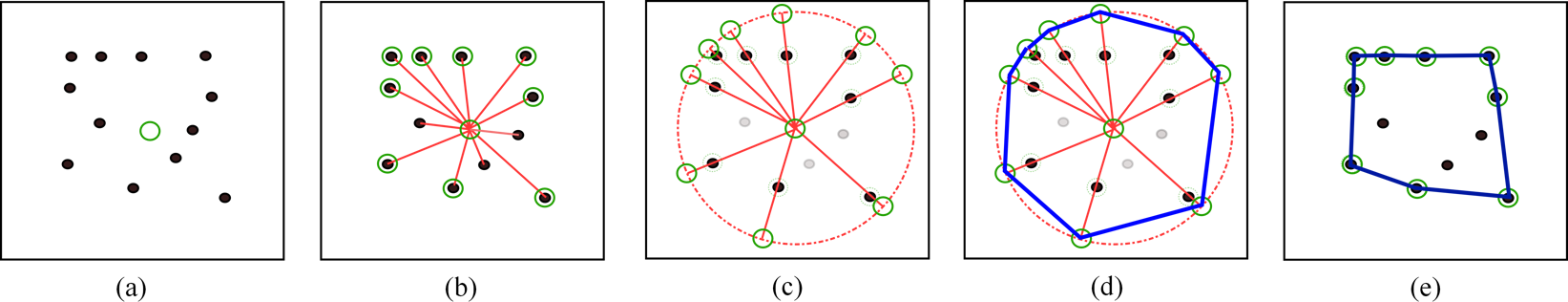}
 \caption[Concept]{Iterative approach for dealing with coplanar triangles - (a) Calculate the shape centroid, (b) determine outer surface points using support mapping (i.e., the normal from the centroid to each point), (c) project surface points onto a unit sphere (i.e., centroid to surface point fixed length), (d) use the projected surface points to find the interconnected convex hull triangles, and (e) the indexes for each of the projected surface points are the convex hull.}
}



\hypersetup{pdfinfo={
   Author		= {\myauthor},
   Title		= {\mytitle \mysubtitle},
   Subject 		= {\mytitle \mysubtitle},
   CreationDate = {D:20130530195600},
   Keywords 	= {\mykeywords},
}}

\pdfinfo{
   /Author (\myauthor)
   /Title  (\mytitle \mysubtitle)
   /Subject (\mytitle \mysubtitle)
   /CreationDate (D:20130530195600)
   /Keywords (\mykeywords)
}




\maketitle

\begin{abstract}
Writing an uncomplicated, robust, and scalable three-dimensional convex hull algorithm is challenging and problematic.  
This includes, coplanar and collinear issues, numerical accuracy, performance, and complexity trade-offs.
While there are a number of methods available for finding the convex hull based on geometric calculations, such as, the distance between points, but do not address the technical challenges when implementing a usable solution (e.g., numerical issues and degenerate cloud points).
We explain some common algorithm pitfalls and engineering modifications to overcome and solve these limitations.  We present a novel iterative method using support mapping and surface projection to create an uncomplicated and robust 2d and 3d convex hull algorithm. 

\end{abstract}

\section{Introduction}

\paragraph{Problem}
Convex hull algorithms are an essential multi-discipline technique important to several fields, such as, computer graphics, pattern recognition, medical analysis and design automation \cite{barber1996quickhull,bentley1978divide,dgregorius,graham1983finding}.  
While multiple approaches are available (e.g., gift wrapping \cite{chand1970algorithm} and divide-and-conquer \cite{preparata1977convex}), writing a stable and robust three dimensional implementation is difficult and challenging.
Since implementing an algorithm in two dimensions may be easy, but not so for three dimensions \cite{avis1995good}.
We survey a number of techniques and address common problems with convex hull algorithms in practice.  Since highly complex and degenerate vertices arise in practice making it difficult to generate a reliable convex hull painlessly.   
We present a novel method of projecting the convex surface points onto a spherical boundary to remedy numerical sensitivity that produces a simple and reliable solution for both two and three dimensional problems. 



\paragraph{Motivation}
The mathematics for generating a convex hull from a set of points is well defined, yet there is no de factor standard algorithm or implementation.
A number of innovative and interesting concepts have been published that solve the problem, yet the implementation of a robust 3D convex hull algorithm is paved with technical challenges \cite{avis1995good}.
The emphasis of this article is on a novel straightforward algorithm to produce a stable implementation that can be applied in both 2D and 3D easily.

\paragraph{Challenges}

While a number of innovative and original convex hull algorithms have been presented, they do not address practical short-comings.  Typically, the algorithms are presented in the context of simple test cases, such as, two-dimensions, to explain the working concept.  We focus on the engineering enhancement necessary for a real-world implementation.  We create a variety of test cases to evaluate the success and failure of the implementation based upon specific problems (see Figure \ref{fig:issues}).  Our solution does not aim for an optimal answer, instead we focus on an uncomplicated method that will generate an accurate convex hull reliably.



\paragraph{Previously}
As we would expect, the convex hull problem has been well studied over the past few decades and resulted in a number of solutions (see Figure \ref{fig:timeline}).
There are two main class of algorithm for solving convex hull problems: insertion algorithms and pivoting algorithms \cite{avis1995good}.  With our implementation sitting within the insertion regime. 
An example of a successful insertion method, is the qhull \cite{barber1996quickhull,qhull} algorithm, which solves precision issues caused by coplanar points by merging facets.  This includes, merging a point into a coplanar facet, merging concave facets, merging duplicate ridges, and merging flipped facets.
Similarly, a pivoting algorithm implementation used by the open source Bullet Physics Engine \cite{coumans2012bullet}, is able to generate reliable convex hulls for collision detection problems.
The Bullet convex hull implementation is based on Preparata and Hong \cite{preparata1977convex} method. This has a time complexity of $O(n log n)$.  Furthermore, to make the algorithm less sensitive to rounding errors, all computations are done with integer math.  The algorithm handles degenerate cases, including arbitrary flat and parallel faces.  While our method is based upon \cite{eddy1977new} (QuickHull), but reduces the computation of extremely sensitive collinear and coplanar issues by identifying outer hull points and projecting them onto a common spherical surface.

\figuremacroF
{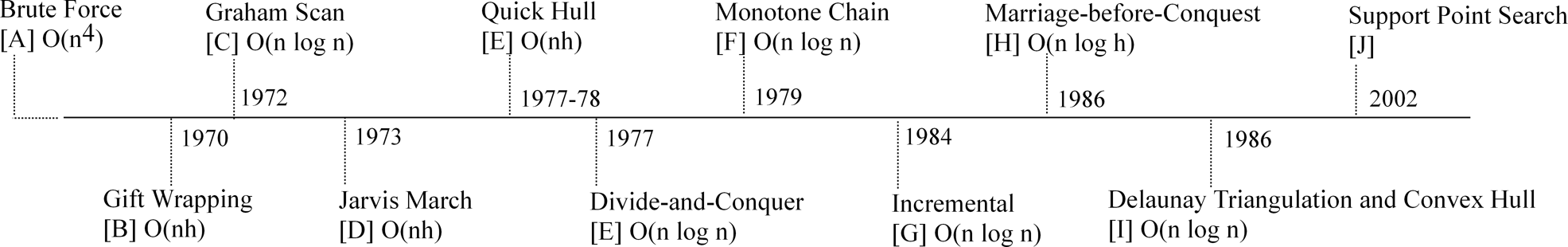}
{Timeline}
{Let $n$ is the number of points in the input set, and $h$ is the number of vertices on the output hull.
[A] Brute Force 
[B]  \cite{chand1970algorithm} 
[C]  \cite{graham1972efficient} 
[D]  \cite{jarvis1973identification} 
[E]  \cite{eddy1977new,bykat1978convex} 
[F]  \cite{preparata1977convex} 
[G]  \cite{andrew1979another} 
[H]  \cite{kallay1984complexity} 
[I]  \cite{kirkpatrick1986ultimate} 
[J] EPA - Expanding Polytope Algorithm \cite{kim2002deep}
[K] \cite{maus1984delaunay}
}
{1.0}

\paragraph{Contribution}
Our main contributions include a definition for a novel convex hull algorithm with less numerical sensitivity and the ability to deal with collinear and coplanar vertices.
The algorithm presented in this article offers a number of desirable benefits:

\begin{itemize}
\item[\checkmark] Easy to implement (i.e., both 2D and 3D)
\item[\checkmark] Terminating condition
\item[\checkmark] Can be used with a tolerance in the expansion step for automatic simplification
\item[\checkmark] Handle `co-planar' and `degenerate' input data
\item[\checkmark] The output mesh is built entirely from the input vertices
\item[\checkmark] Can be applied to real-world complex models, not just point clouds
\end{itemize}

Our approach is the engineering enhancement of mapping the surface points onto a uniform sphere to solve a number of technical shortcomings (e.g., user intervention (tweaking) and numerical issues common with coplanar and collinear faces).


\section{Related Work}
The convex hull problem has received considerable attention in computational geometry \cite{dgregorius,graham1983finding,kim2002deep,maus1984delaunay}.
Computing a convex hull (or just a ``hull'') is one of the first sophisticated geometry algorithms, and there are many variations of it. The most common form of this algorithm involves determining the smallest convex set (called the ``convex hull'') containing a discrete set of points. This algorithm also applies to a polygon, or just any set of line segments, whose hull is the same as the hull of its vertex point set. There are numerous applications for convex hulls, for instance, collision avoidance, hidden object determination, and shape analysis. 

The most popular hull algorithms are the ``Graham scan'' algorithm \cite{graham1972efficient} and the ``divide-and-conquer'' algorithm \cite{preparata1977convex}. Implementations of both these algorithms are readily available (see \cite{o1998computational}).  Both are $O(n log n)$ time algorithms, but the Graham has a low run-time computation overhead in 2D. However, the Graham algorithm does not generalize to 3D and higher dimensions, whereas the divide-and-conquer algorithm has a natural extension.
Figure \ref{fig:timeline} shows the time-line of hull algorithms. 

Our work is based around the insertion algorithm concept.  Where an initial convex hull approximation is created (i.e., a starting tetrahedron for 3D).  We use the support mapping (e.g., see Expanding Polytope Algorithm (EPA) \cite{kim2002deep}) and Quickhull methodology to iteratively grow and encapsulate all the points to form a convex hull.

\figuremacroFb
{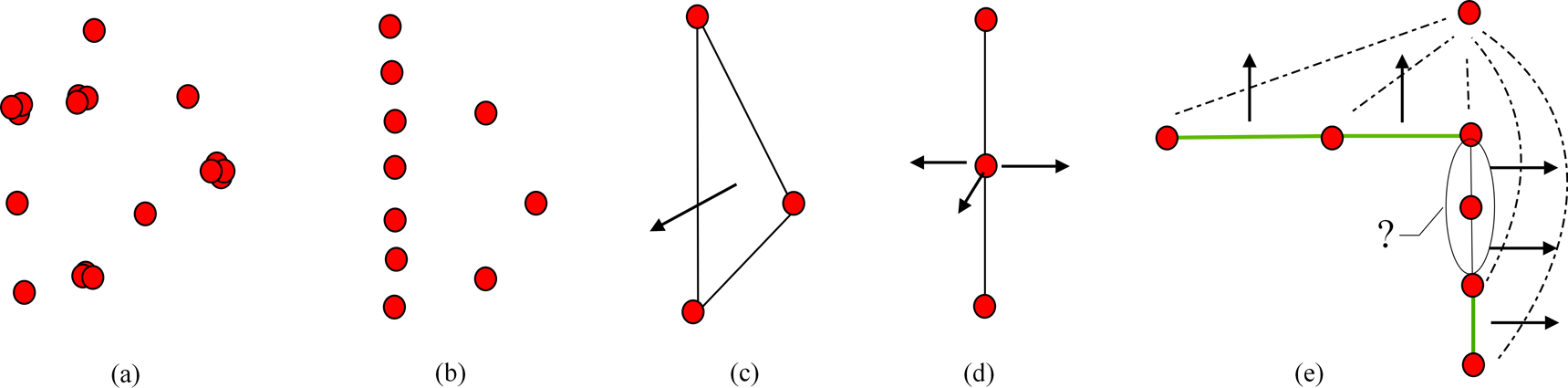}
{Implementation Considerations}
{Example problems, such as, (a) multiple overlapping points, (b) co-planar points, (c) valid triangle, (d) degenerate triangle (i.e., long and thin causes issues when calculating face normals - sliver-shaped triangles), and (e) coplanar points - causing incorrect selection of faces for an expanding convex polytope (i.e., overlapping triangles resulting in concave errors as shown in Figure \ref{fig:fail2}).}
{1.0}

\paragraph{Which method is best?} It depends on what you want, for example, do you want a 2D or a 3D solution?  Are you concerned with run-time speeds or numerical stability and accuracy?

\begin{itemize}
\item[-] Parallelizable
\item[-] Memory Overhead
\item[-] Complexity
\item[-] Computational Speed
\item[-] Robustness and Numerical Sensitivity
\item[-] Number of Dimensions (2D or 3D)
\end{itemize}

\section{Background}
\paragraph{Convex Hulls}
A convex hull means the smallest convex region which encloses a specified group of points.
Technically, it is the smallest convex set containing the points, and can be visualized as a rubber band which wraps around the `outside' points (i.e., all other points must lie within this rubber band \cite{barber1996quickhull}).
A convex hull is different for dissimilar objects because it depends upon the feature point of every object.  For a detailed explanation of Convex Geometry, see Joseph O'Rourke \cite{o1998computational}.
A convex hull of a set is unique (upto co-linearities).
Our method of surface mapping and projecting the points onto a sphere reduces sensitivity and co-linearity ambiguities.


\paragraph{Support Mapping}
Support mapping is often used in physics and collision detection \cite{kim2002deep}.  The support mapping for a cloud of points given a direction is the point that is farthest in the direction - which simply means finding the point with the maximum dot product (i.e., dot(direction, point). The supporting point in any direction is guaranteed to be on the surface of the convex hull cloud of points.  We exploit this concept in our algorithm to efficiently determine the surface points.

\paragraph{No New Vertices}
We have an array of points and want to find how they can be connected using triangles to form a convex hull.  No extra points are added.  The vertices are numbered from $1$ to $N$, with each triangle formed by an array of three indices into the vertex array. This is to avoid any numerical drifting.  We work with triangle and vertex indices and do not generate any new points.

\figuremacroW
{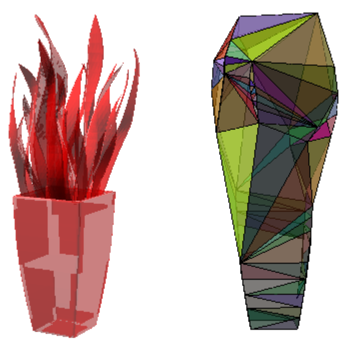}
{Complex 3D Model}
{Complex model of a plant (19370 faces) mesh and generated convex shell (198 faces).}
{0.7}

\paragraph{Centroid}
A bounded convex polyhedron is called a polytope.
The centroid of a convex polytope as the centroid of its vertices is given by Equation \ref{eq:centroid}.

\vspace{-12pt}
\begin{equation} \label{eq:centroid}
\frac{1}{n} \sum_{i=1}^{n} p_i
\end{equation}

\noindent where the centroid is composed of a set of points ${p_1,...p_n}$.  Note, the centroid will be contained within the relative interior of the convex hull.

\paragraph{Algorithm Overview}
Our algorithm adopts the well-known QuickHull approach but with additional pre-phase culling and re-mapping of the vertices. It starts by calculating the centroid and performing a support mapping phase to strip inner vertices, the final vertices are then mapped onto a spherical surface using the centroid.  From the set of points we use four points to generate a
tetrahedron (note - due to support mapping phase, we will not need to discards any internal points as the hull grows).  It then iteratively refines the faces of the polyhedron by adding external points, and redistributes the remaining points associated with each face among its children faces. The refinement of a face is performed by selecting the furthest point from its associated points and generating three children triangles.  We do not need to worry about concave edge swapping or removing concave vertices.

\section{Our Iterative Method}
The convex hull algorithm presented in this paper focuses on 3D cases.  We are mainly interested in computing convex hulls that are able to solve unforeseen problems for arbitrary clouds of points, which can contain degenerate data, in addition to model scene geometry, such as, complex 3D geometric models (e.g., see Figure \ref{fig:guntest}, Figure \ref{fig:minkowskicubes}, and Figure \ref{fig:cattest}).

A preliminary stage is run to strip out and prepair the points.
The algorithm starts with a set of points.
We remove duplicate points (i.e., points within a predefined tolerance).
We calculate the centroid of the set of points).
We calculate the normal from the centroid to each point and find the support vertex, and add it to a list.
The list will contain a set of points which sit on the convex hull surface.
Using the centroid, we project each surface point onto a sphere (i.e., with the centroid the centre of the sphere).

The iterative stage starts by taking the first four vertices and connecting them to form the smallest possible starting closed mesh (i.e., a tetrahedron).
After we have set-up the tetrahedron, we iteratively grow the convex hull to encapsulate the rest of the surface points and create the convex hull.
We grow the convex hull by going through each of the surface points, and selecting the triangle surfaces that are visible to the point.
The visible faces are removed and a new set of faces are added using the edges are are not shared and the new surface point.  The key stages are given in Algorithm \ref{alg:myalgorithm}.

\begin{itemize}
\item \textbf{Point inside or outside a convex shape} - we can easily determine if a point is inside a convex hull by iterating over all the faces and checking if the point is on the inside of the plane (i.e., dot product).  This can be useful for automatically checking if the algorithm failed when developing the implementation.
\item \textbf{Faces that a point can see} - we find all the faces that are visible to a point by taking the dot product of the face normal and the point (i.e., front facing if $n \bullet p > n \bullet v$, where $n$ is the face normal, $v$ is a face vertex, and $p$ is the test point). 
\item \textbf{Extract edges from a set of faces} - we have an array of edges from all the found triangles, any edges that are shared (i.e., $count > 1$), are thrown away.  Then the remaining edges are used to create new triangles (i.e., edge and the new point).
\end{itemize}

\paragraph{Dynamics}
Our method is able to handle unknown sets of points. In addition, due our algorithms iterative nature, we are able to handle changing concave hulls, where points can be added or deleted on-the-fly.  Our convex hull algorithm is easily able to update the mesh after each insertion/deletion operation.



\paragraph{Modifications}
The algorithm is flexible and can be modified to approach the problem in different ways.  For example, instead of iteratively selecting each vertex in the list as we do in our implementation, we could exploit the support mapping concept further by iterating over each face and select the point furthest from the face to iteratively grow the convex polytope.
Our algorithm relies on a simple local geometric point-plane test to determine the position of a point with respect to a plane, which is
used to pick the triangles to merge the point with.   As we have already culled inner points with the support mapping phase.
All the points after the support mapping phase are used to create the convex hull surface.  Projecting the points onto a sphere reduces co-planar and co-linear issues.  Due to the approach the method does not require any swap operations to resolve fold-overs and self-intersections which can complicate the point-plane test and disturb their locality \cite{stein2012cudahull}.

\paragraph{Optimisation}
We did not focus on any optimisations, but on a novel solution for providing a robust and easy to implement method that resolves common issues (i.e., reducing coplanar and collinear points).  However, the timing results for our implementation are given in Figure \ref{fig:timing} to shows the relationship between the number of vertices and the elapsed time.  
Our iterative algorithm has a time complexity $O(n\; log \; n)$ since it is built upon the concept presented by Clarkson and Shor \cite{clarkson1989applications} which iteratively adds external point to extend the convex polyhedron until the remaining set of points becomes empty.

\begin{algorithm}[h]
 \caption{Iterative 3D Surface Mapping onto a Sphere Convex Hull Algorithm}
 \label{alg:myalgorithm}
 \KwData{Array Points}
 \KwResult{Triangle array representing the convex hull for the set of Points}
 Remove duplicate points (i.e., within pre-defined distance tolerance)\\
 Calculate Centroid\\
 Empty array spherepoints\\
 \For{i=1 to num points}
 {
	 Normal = Centroid to Points(i)\\
	 \vspace{2pt}
	 supportpoint = FindSupportPoint(Normal, Points)\\
	 \vspace{2pt}
	 Project surfacepoint onto sphere surface\\
	 \vspace{2pt}
	 Add supportpoint to spherepoints\\
 }
 
 Array hull\\
 Construct a tetrahedron using spherepoints (e.g., first four points points) add to array hull\\
 \For{i=4 to num spherepoints}
 {
	Find all triangles that are visible to spherepoint(i) (i.e., front of the triangle) \label{alg:line0}\\

	\vspace{2pt}
	
	Remove found triangles from array hull\\
	
	\vspace{2pt}
	Find non-shared edges for the removed triangles
	
	\vspace{2pt}
	Add new triangles using spherepoint(i) and non-shared edges
 }

\end{algorithm}

\figuremacroFb
{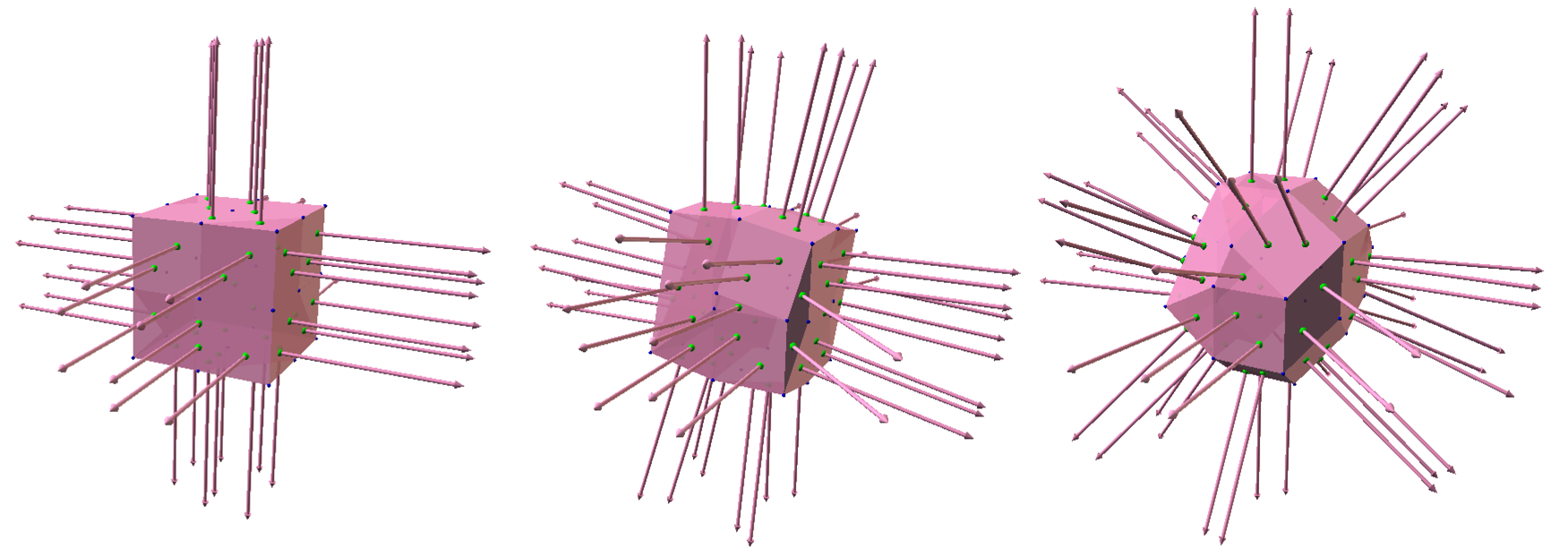}
{Minkowski Sum}
{Summing two cube models using the Minkowski Sum to evaluate the Convex Hull implementation in a practical situation.  Visually drawing each triangle normal allows us to easily identify the Convex Hull Mesh. (One cube is rotated while the second remains fixed for the Mikowski sum test cases shown in the illustration).}
{1.0}



\section{Experimental Results}
We implemented the algorithm using floating point precision in C++ within Visual Studio 2013 and Windows-7.
We evaluated our implementation using various test scenarios:
\begin{itemize}
\item Various complex 3D models (e.g., gun, rabbit, teapot) - Figure \ref{fig:cattest}, \ref{fig:guntest}
\item Procedural test case (e.g., Minkowski shape) - Figure \ref{fig:minkowskicubes} - this has the added advantage of generating a wide variety of point data (e.g., degenerate cases that may not naturally occur in preloaded mesh models)
\item Random point clusters - i.e., to provide approximate performance metrics for the computational cost versus the number of points - Figure \ref{fig:timing}
\end{itemize}

We also emphasis failure cases, as shown in Figure \ref{fig:fail1}, Figure \ref{fig:fail2}, and Figure \ref{fig:fail3}, which are caused when we do not include the additional spherical surface projection phase to reduce co-linear and coplanar issues.  A point to note, is our approach reduces accuracy constraints (i.e., numerical sensitivity) for different curved surfaces by projecting the points onto a common spherical lattice.
While we can use support mapping to extrapolate the surface vertices, incorrectly expanding the initial tetrahedron, due to numerical sensitivity and coplanar triangles, can produce mesh that engulfs the surface vertices but be concave, as shown in Figure \ref{fig:fail2}.

The randomly generate point set tests of different sizes ranged
from 500 to 10000 points. For each size, we generate 5 different datasets and average their run time. Figure \ref{fig:timing} shows the run time details and shows our algorithm is $O(n\;log\;n)$. 

\figuremacroW
{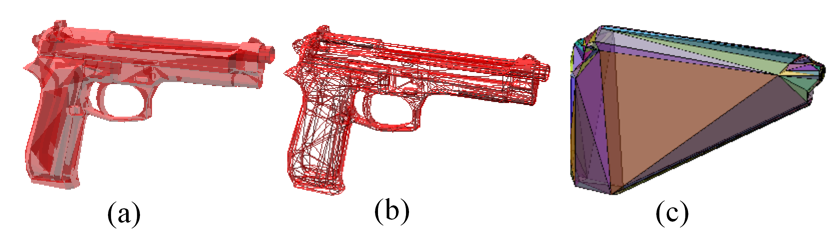}
{Complex 3D Model}
{(a) Complex model of a gun (6174 points or 2058 faces), (b) wireframe view, and (c) generated convex shell (342 points or 114 faces).}
{1.0}


\figuremacroW
{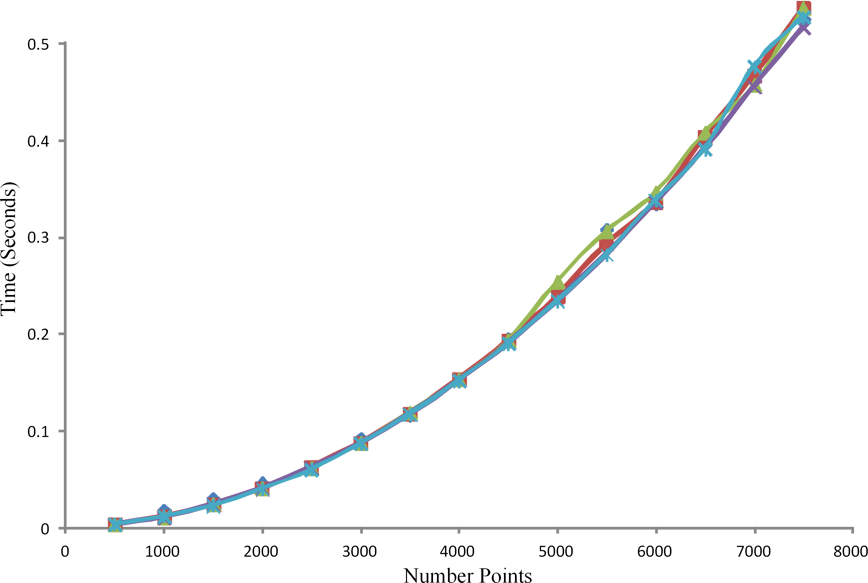}
{Random 3D Cloud of Points}
{Approximate performance metrics measuring the time to generate a convex hull given different sets of randomly generated 3D points (i.e., from 500 to 10000 vertices).}
{1.0}

\section{Discussion}
A number of factors come into question, such as, computational speed and robustness.
On the surface, a convex algorithm may appear elegant and straightforward but can be difficult to implement well.  For example, for 3D point clouds, the solution can hit numerical issues for small hulls and computational bottlenecks for large numbers of points (10,000 or more vertices).
While fixed-point integer mathematics may help improve robustness, using real-numbers with limited accuracy (e.g., floats or doubles) makes the algorithm much faster but at a cost (e.g., stability and accuracy).
We presented a novel generic method that works effectively with real-numbers that is able to deal with coplanar and collinear surfaces without any complex engineering enhancements or user interventions.

Our method works because a convex hull is defined by its vertices.
The advantage of our method over other approaches is the surface vertices have already been found
during the support mapping phase without overhead.  In terms of complexity, it has the benefit of not having to do vertex culling when generating the convex surface.  Additionally, when combined with sphere mapping, it reduces the need to perform swap operation to fix the convexity due to concave and other edges.

\figuremacroW
{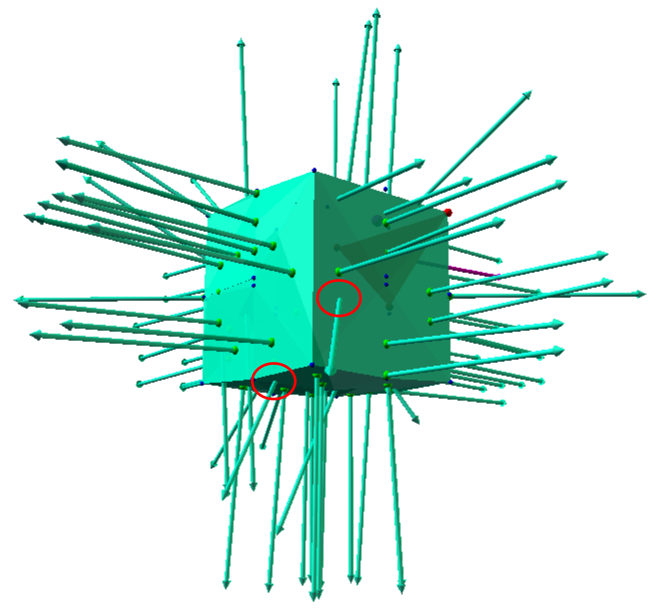}
{Failure Case 1}
{Caution needs to be taken - while an implementation may seem correct on the surface - degenerate and troublesome triangles may be inserted within the convex hull (we can easily make out the troublesome inner triangles by seeing the normals).}
{1.0}

\figuremacroW
{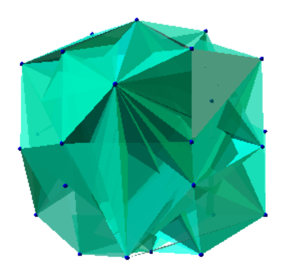}
{Failure Case 2}
{Incorrectly removing and adding triangles iteratively may engulf all the surface vertices but produce a concave mesh (i.e., non-convex).}
{0.8}

\figuremacroW
{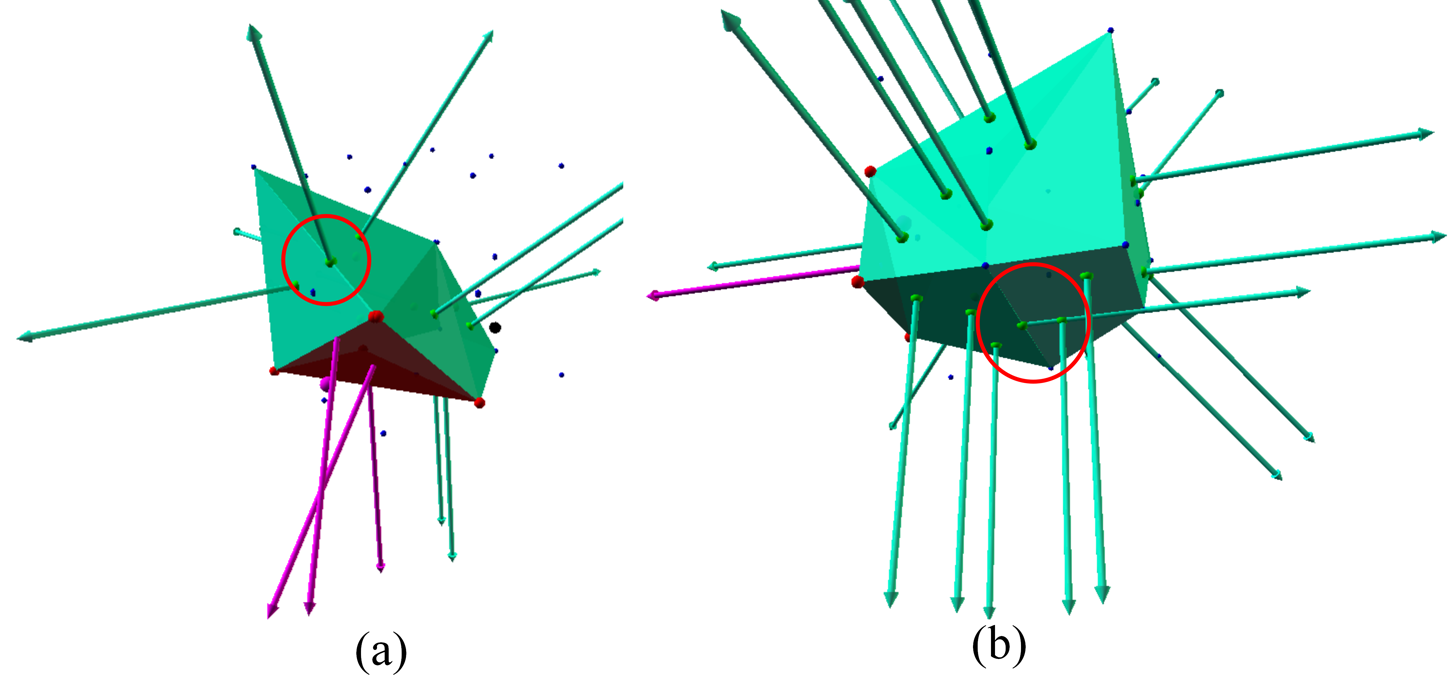}
{Failure Case 3}
{Degenerate triangles (i.e., long thin triangles) can cause issues with the iterative evolution of the convex hull (e.g., incorrect normals).  (a) long thin triangle, and (b) a degenerate triangle with a normal that is parallel to the surface.}
{1.0}

\figuremacroW
{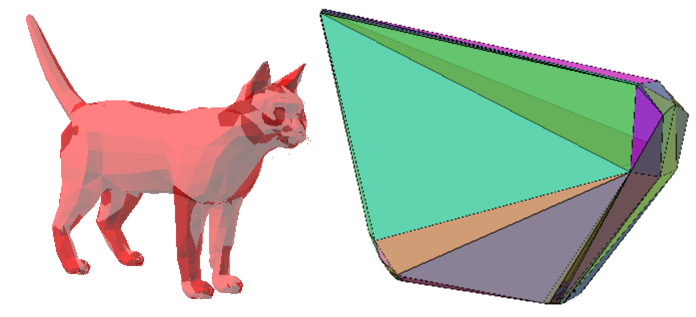}
{Complex 3D Model}
{(a) Complex model of a cat (11862 points or 3954 faces), and (b) generated convex shell (162 points or 54 faces).}
{1.0}

%

\section*{Acknowledgements}
A special thanks to reviewers for taking time to review this article and provide insightful comments and suggestions to help to improve the quality of this article.


\bibliographystyle{acmsiggraph}



\let\oldthebibliography=\thebibliography
\let\endoldthebibliography=\endthebibliography

\renewenvironment{thebibliography}[1]{%
    \begin{oldthebibliography}{#1}%
      \setlength{\parskip}{0ex}%
      \setlength{\itemsep}{0.5ex}%
  }%
  {%
    \end{oldthebibliography}%
  }


\bibliography{paper}

\def\url#1{}
\begin{thebibliography}{\protect\citename{Graham and Frances~Yao }1983}

\bibitem[\protect\citename{Andrew }1979]{andrew1979another}
{\sc Andrew, A.~M.}
\newblock 1979.
\newblock Another efficient algorithm for convex hulls in two dimensions.
\newblock {\em Information Processing Letters 9}, 5, 216--219.

\bibitem[\protect\citename{Avis and Bremner }1995]{avis1995good}
{\sc Avis, D., and Bremner, D.}
\newblock 1995.
\newblock How good are convex hull algorithms?
\newblock In {\em Proceedings of the eleventh annual symposium on Computational
  geometry}, ACM, 20--28.

\bibitem[\protect\citename{Barber et~al\mbox{.} }1996]{barber1996quickhull}
{\sc Barber, C.~B., Dobkin, D.~P., and Huhdanpaa, H.}
\newblock 1996.
\newblock The quickhull algorithm for convex hulls.
\newblock {\em ACM Transactions on Mathematical Software (TOMS) 22}, 4,
  469--483.

\bibitem[\protect\citename{Barber }1995-2012]{qhull}
{\sc Barber, C.}, 1995-2012.
\newblock http://www.qhull.org/.

\bibitem[\protect\citename{Bentley and Shamos }1978]{bentley1978divide}
{\sc Bentley, J.~L., and Shamos, M.~I.}
\newblock 1978.
\newblock Divide and conquer for linear expected time.
\newblock {\em Information Processing Letters 7}, 2, 87--91.

\bibitem[\protect\citename{Bykat }1978]{bykat1978convex}
{\sc Bykat, A.}
\newblock 1978.
\newblock Convex hull of a finite set of points in two dimensions.
\newblock {\em Information Processing Letters 7}, 6, 296--298.

\bibitem[\protect\citename{Chand and Kapur }1970]{chand1970algorithm}
{\sc Chand, D.~R., and Kapur, S.~S.}
\newblock 1970.
\newblock An algorithm for convex polytopes.
\newblock {\em Journal of the ACM (JACM) 17}, 1, 78--86.

\bibitem[\protect\citename{Clarkson and Shor }1989]{clarkson1989applications}
{\sc Clarkson, K.~L., and Shor, P.~W.}
\newblock 1989.
\newblock Applications of random sampling in computational geometry, ii.
\newblock {\em Discrete \& Computational Geometry 4}, 1, 387--421.

\bibitem[\protect\citename{Coumans }2012]{coumans2012bullet}
{\sc Coumans, E.}, 2012.
\newblock Bullet physic sdk manual.

\bibitem[\protect\citename{Eddy }1977]{eddy1977new}
{\sc Eddy, W.~F.}
\newblock 1977.
\newblock A new convex hull algorithm for planar sets.
\newblock {\em ACM Transactions on Mathematical Software (TOMS) 3}, 4,
  398--403.

\bibitem[\protect\citename{Graham and Frances~Yao }1983]{graham1983finding}
{\sc Graham, R.~L., and Frances~Yao, F.}
\newblock 1983.
\newblock Finding the convex hull of a simple polygon.
\newblock {\em Journal of Algorithms 4}, 4, 324--331.

\bibitem[\protect\citename{Graham }1972]{graham1972efficient}
{\sc Graham, R.~L.}
\newblock 1972.
\newblock An efficient algorith for determining the convex hull of a finite
  planar set.
\newblock {\em Information processing letters 1}, 4, 132--133.

\bibitem[\protect\citename{Gregorius }2014]{dgregorius}
{\sc Gregorius, D.}
\newblock 2014.
\newblock Implementing quickhull.
\newblock {\em Game Developers Conference (Valve Software) in San Francisco\/}.

\bibitem[\protect\citename{Jarvis }1973]{jarvis1973identification}
{\sc Jarvis, R.~A.}
\newblock 1973.
\newblock On the identification of the convex hull of a finite set of points in
  the plane.
\newblock {\em Information Processing Letters 2}, 1, 18--21.

\bibitem[\protect\citename{Kallay }1984]{kallay1984complexity}
{\sc Kallay, M.}
\newblock 1984.
\newblock The complexity of incremental convex hull algorithms in rd.
\newblock {\em Information Processing Letters 19}, 4, 197.

\bibitem[\protect\citename{Kim et~al\mbox{.} }2002]{kim2002deep}
{\sc Kim, Y.~J., Lin, M.~C., and Manocha, D.}
\newblock 2002.
\newblock Deep: Dual-space expansion for estimating penetration depth between
  convex polytopes.
\newblock In {\em Robotics and Automation, 2002. Proceedings. ICRA'02. IEEE
  International Conference on}, vol.~1, IEEE, 921--926.

\bibitem[\protect\citename{Kirkpatrick and Seidel
  }1986]{kirkpatrick1986ultimate}
{\sc Kirkpatrick, D.~G., and Seidel, R.}
\newblock 1986.
\newblock The ultimate planar convex hull algorithm?
\newblock {\em SIAM journal on computing 15}, 1, 287--299.

\bibitem[\protect\citename{Maus }1984]{maus1984delaunay}
{\sc Maus, A.}
\newblock 1984.
\newblock Delaunay triangulation and the convex hull ofn points in expected
  linear time.
\newblock {\em BIT Numerical Mathematics 24}, 2, 151--163.

\bibitem[\protect\citename{O'Rourke }1998]{o1998computational}
{\sc O'Rourke, J.}
\newblock 1998.
\newblock {\em Computational geometry in C}.
\newblock Cambridge university press.

\bibitem[\protect\citename{Preparata and Hong }1977]{preparata1977convex}
{\sc Preparata, F.~P., and Hong, S.~J.}
\newblock 1977.
\newblock Convex hulls of finite sets of points in two and three dimensions.
\newblock {\em Communications of the ACM 20}, 2, 87--93.

\bibitem[\protect\citename{Stein et~al\mbox{.} }2012]{stein2012cudahull}
{\sc Stein, A., Geva, E., and El-Sana, J.}
\newblock 2012.
\newblock Cudahull: Fast parallel 3d convex hull on the gpu.
\newblock {\em Computers \& Graphics 36}, 4, 265--271.

\end{thebibliography}

\end{document}